\documentclass[a4paper,natbib]{apa6}

\usepackage{hyperref}

\usepackage[english]{babel}
\usepackage[utf8x]{inputenc}
\usepackage{amsmath}
\usepackage{graphicx}
\usepackage[colorinlistoftodos]{todonotes}

\begin{document}

\title{Contextuality in Human Decision Making in the Presence of Direct Influences: A Comment on Basieva et al. (2019).}
\shorttitle{Comment on Basieva et al. (2019)}

\twoauthors{James M Yearsley}{Jonathan J Halliwell}
\twoaffiliations{Department of Psychology,\\ City, University of London,\\ EC1V 0HB, UK}{Blackett Laboratory,\\ Imperial College,\\ London SW7 2AZ, UK}

\abstract
{In a recent paper \cite{Basieva2019} presented a series of experiments which they claimed show evidence for contextuality in human judgments. This was based on a set of modified Bell-like inequalities designed to rule out effects caused by signalling. 
In this comment we show that it is, however, possible to construct a non-contextual model which explains the experimental data via {\it direct influences}, which we take to mean that a measurement outcome has a (model-specific) causal dependence on other measurements.
We trace the apparent inconsistency to a definition of signalling which does not account for all possible forms of direct influence. 
Further, we cast doubt on the idea that {\em any} experimental data in psychology could provide conclusive evidence for contextuality
beyond that explainable by direct influence.}

\maketitle

\section{Introduction}

As part of the broader quantum cognition program a number of researchers have considered whether there is evidence for violation of contextual inequalities in psychology (e.g.,  \cite{Aerts2013},  \cite{Asano2014}, \cite{Bruza2015},  \cite{Bruza2015b}). Such inequalities are derived on the assumption that there exist hidden joint preference or probability states for the psychological observables being measured. This is, loosely, equivalent to assuming that judgment processes giving rise to choices between different options operate independently, which is an important constraint on the processes underlying human decision making.

One complicating factor is that it is hard to rule out the possibility of direct influence between measurements, which can mimic the effect of true contextuality. In other words, contextuality means the outcome of a judgment about observable A has an apparent and unexpected dependence
on what else is being measured, but that can also occur if the outcome of the other measurements directly influence A.

The absence of such direct influences must be justified or explicitly tested in any particular application. In physics such influences can sometimes be ruled out by reference to some physical principle, but nothing equivalent in psychology can be used to rule out direct influences a priori. The challenge of quantifying exactly when violations of contextual inequalities can be accounted for by direct influences, and when they can only be explained by genuine contextuality, was taken up by \cite{DK} who derived modified inequalities which, they claim, allow the identification of true contextuality not explainable by 
``signalling'', a specific form of direct influence which we will explain below.

In a series of papers Dzhafarov and collaborators re-analyzed existing experimental claims of contextuality in psychology and concluded they could all be explained by signalling (\cite{Dzhafarov2015}, \cite{Dzhafarov2016}). However in an elegant paper \cite{Cervantes2018} presented an experiment which did satisfy their modified contextual inequality, and in a recent paper \cite{Basieva2019} followed this up with series of experiments, the majority of which produced data demonstrating genuine contextuality, ie over and above that explicable by signalling, according to Dzhafarov and Kujala's modified inequalities.

In this comment we explain why we nevertheless do not believe that the results presented by \cite{Basieva2019} provide conclusive
 evidence for contextuality in human decision making.  We only consider in detail the form of experiment conducted by \cite{Basieva2019} and \cite{Cervantes2018}, but we use the insight gained to question whether contextuality could ever be observable in human decision making.

\section{Outline of \cite{Basieva2019}}

Consider one of the experiments in \cite{Basieva2019}; 

``Alice wishes to order a two-course meal. For each course she can choose a high-calorie option (indicated by H) or a low-calorie option (indicated by L). Alice does not want both courses to be high-calorie nor does she want both of them to be low-calorie.''

Each participant was given two out of three courses (Starter, Main, Dessert) to choose from. Clearly participants should select options so that while the calorie content of, eg, the starter is undetermined, it is anti-correlated with the calorie content of the main (indeed this restriction was forced on participants in the experiment.) We can easily see why \cite{Basieva2019} expect to see evidence of contextuality - the underlying probability distribution for all three courses needs to have three binary anti-correlated variables, which is impossible. In other words, the calorie content of at least two of the dishes has to match, since there are three courses and only two calorie options, but then one cannot have the three choices anti-correlating.

The specific inequality \cite{Basieva2019} test  is an example of modified Bell-type inequalities derived by Dzhafarov and Kujala (2015).
They use the contextuality by default (CbD) approach, in which a set of random variables $R_n$ each taking values $\pm 1 $ may take different values in different measurement contexts. So $R_n^m$ denotes the value of $R_n$ in measurement context labeled by $m$. In the above example $n,m=1,2,3$ and context $1$ might be
the condition where participants were asked to choose options for Starter and Main. In the CbD approach, direct influence is defined in terms of the degree to which $R_n^m$ and $R_n^k$ may be different.

The modified Bell-type inequalities consist of standard Bell-type inequalities but modified by terms of the form
\begin{equation}
   \Delta = |E[R^1_1]-E[R^3_1]|+|E[R^1_2]-E[R^2_2]|+|E[R^2_3]-E[R^3_3]|,
\label{eq1}
\end{equation}
where $E[\cdot]$ denotes an expectation value. The quantity $\Delta$ is held to be a measure of signalling in this situation, which is thus seen to be defined in terms of an {\it average} of the degree of direct influence, since it involves only terms of the form $E[R_n^m - R_n^k]$.
Such modified inequalities permit the identification of contextuality beyond that explainable by signalling.
In the specific model considered by \cite{Basieva2019} the modified Bell-type inequality has the simple form
\begin{equation}
\Delta < 2,
\label{eq1.5}
\end{equation}
here written in a notational form opposite to the traditional form of the Bell inequalities so contextuality is deemed to be present if this equation is satisfied.
What  \cite{Basieva2019}
did was to show this inequality was satisfied by data collected in a number of different experiments which were variants of the one outlined above. 

However it is intuitively obvious how participants could solve the problem set by \cite{Basieva2019}; given two courses to select, eg starter and main, choose the calorie content of the first one randomly, then make the opposite choice for the second course. This complies with the instructions and is ``non-contextual'' in a colloquial sense, since it makes no reference to measurement contexts. However it does not sit well with the idea of a pre-existing preference, since one of the judgments is made deterministically based on the other, with no reference to existing preferences. We therefore need to apply a more precise measure of contextuality. Also, as described, this strategy has the feature that it will tend to produce equal preferences for each course, which is not what was observed in \cite{Basieva2019}. So we need to establish that this heuristic can generalise to cases where the preferences are not equal.

\section{A Possible Non-Contextual Account?}

If the variables of interest are thought to be the same in different contexts (e.g. dish choice for main in the context of starter, dish choice for main in the context of dessert)
non-contextuality means that there exists a joint probability matching the set of marginal probabilities characterizing the data. Contextuality is thus defined to be the absence of such a distribution. This definition of contextuality is essentially the same as that frequently employed both in physics (see for example \cite{Abramsky2011}) and in cognitive models in psychology (see for example, \cite{Oaksford2007}). 
If the variables are allowed to take different values in different contexts, then the way this standard notion of non-contextuality is extended in the CbD approach is to require that the variables vary as little as possible across different contexts (Dzhafarov and Kujala, 2015), about which more below.

Let us begin with our idealized version of the experiment: assume  participants solve the problem by choosing the calorie content of the first course randomly, then making the opposite choice for the second course.   The expectation value of any of the variables therefore equals zero, regardless of the context in which it is measured. That means,
\begin{equation}
   \Delta =0,
\end{equation}
so Eq.(\ref{eq1.5}) is satisfied and  \cite{Basieva2019}
would presumably claim genuine contextuality in this case, i.e. contextuality over and above that which could be explained by signalling. 

However it is still possible to write down a probability distribution on the variables $R^1_1, R^1_2, R^2_2, R^2_3, R^3_1, R^3_3$, which has these correlations and expectation values:
\begin{equation}
p(R^1_1, R^1_2, R^2_2, R^2_3, R^3_1, R^3_3)=\frac{1}{64}(1-R^1_1 R^1_2)(1-R^2_2 R^2_3)(1-R^3_1 R^3_3).
\end{equation}
Note $E[R^1_1, R^1_2]=-1$ etc, as required, and $E[R^i_j]=0$ for all variables and contexts. This proves that a particular type of non-contextual account of this idealisation of the \cite{Basieva2019} experiments is possible. The explanation is  a direct influence of $R^1_1$ on $R^1_2$ etc,  such that the value of one random variable in a given context is set equal to minus the value of the other one. This does not conform to the standard notion of a  non-contextual model in the CbD approach since it involves probabilities on variables permitted to take different values in different contexts, but without the requirement of minimal variation across different contexts.
However, it is clearly still of interest since it gives a probabilistic explanation of the data in terms of the action of direct influences.

We note an interesting property of  this probability distribution, which is that it clearly factorises as;
\begin{equation}
p(R^1_1, R^1_2, R^2_2, R^2_3, R^3_1, R^3_3)=p(R^1_1,R^1_2)p(R^2_2,R^2_3)p(R^3_1,R^3_3)\label{propfac}
\end{equation}
One consequence is that correlation functions between the same variable in different contexts are zero, eg $E[R^1_1 R^3_1]=0$.  The reason, in terms of a process account, is that $R^3_1$ is basically set by $R^3_3$, which is an independent random variable. So the effect of the direct influence is to remove correlations between the same variable in different contexts.

This idealisation is interesting, because the fact the expectation values of all individual variables are all zero means the modified contextual inequality of Dzhafarov and Kujala's (2015) reduces to the one in the absence of signalling. In other words, although our account of this experiment involves direct influence between variables measured in the same context, 
it does not involve the weaker notion of signalling. 
This suggests the origin of the discrepancy between the claims in \cite{Cervantes2018} and \cite{Basieva2019}  and our construction of a non-contextual model lies in the definition of signalling used by Dzhafarov and Kujala (2015). We will explore this further below.

Our idealisation of the experiments in \cite{Basieva2019} is informative, but the results they reported had non-zero expectation values for $R^1_1, R^2_2$ and $R^3_3$. We can modify our account to deal with this by taking the joint probability to have the same form as Eq.(\ref{propfac}) above, where now,
\begin{equation}
p(R^m_i,R^m_j)= \frac{1}{4}(1+(R^m_i-R^m_j)E[R^m_i]-R^m_i R^m_j)
\end{equation}
where $E[R^m_m]$ are the measured expectation values.

This obviously has the correct values for the measured expectation values and correlations. It also has the same interpretation, namely that there is a direct influence between, eg $R^1_1$ and $R^1_2$, such that, on measuring the value of $R^1_1$, the value of $R^1_2$ is set to minus this. The correlations between  variables measured in different contexts are no longer zero, however we have $E[R^m_i R^n_i]=E[R^m_i] E[R^n_i]$, so they are still independent. There are no constraints on the $E[R^m_m]$ in order that this construction be valid. 

We have therefore shown by explicitly constructing a joint probability distribution that the experimental results reported in \cite{Basieva2019} can be accounted for by a particular type of non-contextual model which includes direct influences. More precisely, a model is possible in which preferences for the three dish choices are well defined at all times, but there is an explicitly modeled disturbance caused by eliciting a preference which explains the apparently contextual data.

\section{Signalling vs Direct Influence}

The roots of our disagreement with \cite{Basieva2019} lie in the work of Dzhafarov and Kujala (2015) which may be regarded as a generalization of the famous result of  \cite{Fine1982}
who established the conditions under which certain sets of marginal probabilities possess a joint probability distribution. A crucial assumption in Fine's work is that overlapping pairwise marginal probabilities are compatible with each other, a condition referred to as marginal selectivity, which in the present application reduces to a set of simple conditions of the form
\begin{equation}
E[R_i^j] = E[R_i^k], \label{E10}
\end{equation}
in other words, the average values of all variables $R_i^j$ are independent of context.
 Dzhafarov and Kujala (2015) essentially demonstrate how to extend Fine's result to embrace the case in which marginal selectivity fails.

This generalized Fine's theorem leads to the set of Bell-type inequalities mentioned above which Dzhafarov and Kujala (2015) claim to be tests for contextuality in the presence of signalling.
As indicated, they define signalling as a failure of conditions such as Eq.(\ref{E10}), or more generally, by non-zero values of the quantity $\Delta$ defined in Eq.(\ref{eq1}).
Since this definition of signalling corresponds to the {\it average} degree of direct influence, it leaves a residual degree of direct influence which has the power to explain apparent contextuality not explained by signalling.
The presence of direct influence is characterized precisely by non-zero values of probabilities of the form $p(R_i^j \ne R_i^k)$ (i.e, the probabilities that the same variable measured in different contexts gives different results). This probability can be non-zero even when Eq.(\ref{E10}) holds. Indeed this possibility occurs in our model above where direct influence is present trial to trial but averages to zero. 

The difference between signalling and more general direct influence is not very apparent in the approach of Dzhafarov and Kujala (2015) since in their definition of non-contextuality, the underlying joint probability is required to change as little as possible across different contexts. They implement this by requiring that the probabilities of the form $p(R_i^j \ne R_i^k)$ are  
{\it minimized}. The minimum then depends only on terms proportional to $\Delta$, Eq.(\ref{eq1}),  hence notions of signalling and more general direct influence coincide in this situation.

To put all this another way, in order to claim contextuality, it is necessary to show that there is no other possible account of the correlations. In physics it is necessary to go to some lengths to be sure of this. The attitude one needs to adopt is of the ``worst case scenario'', where the direct influence is as hard to detect as possible. Only by ruling out this sort of stubborn direct influence can we be sure that a non-contextual account is impossible. In contrast, focussing on changes to the averages can be thought of as a ``best case scenario'', where the direct influence is as easy to detect as possible. Ruling out changes to the average distributions is necessary, but not sufficient to rule out direct influences, because one could imagine, for example, that the process of measuring A changes the correlation between A and B, but not the averages. This clearly implies a direct influence between A and B, but one which is not detectable from the averages alone.

To give a simple example, imagine we have two coins which are tossed either together or independently and under either circumstance both have probability $1/2$ of coming up heads. Suppose however that the coins always come up the same when tossed together. There is no signalling in the sense defined above, but there is clearly a direct influence trial to trial.

The notion of direct influence beyond the average considered here may however not be readily detectable without more elaborate measurements, a key qualitative difference to the weaker notion based just on signalling, which involves readily measurable quantities. In physics such measurements are not hard to devise and higher order signalling conditions that detect direct influences beyond the average have been proposed, e.g. by \cite{CK2015}.
This could be a lot harder in psychological experiments, which means that a definition of of contextuality phrased only in terms of what can actually be measured is a reasonable one.

In summary, we see that the claims of Dzhafarov \& Kujala (2015) about the presence of contextuality beyond that explainable by signalling hinge on a notion of signalling as average direct influence, a notion weaker than that used in physics (where ``signalling'' is more commonly associated with direct influences more generally). We have found that a particular type of non-contextual model is possible if direct influence beyond the average is taken into account.

\section{Discussion}

The above results raise two interesting questions; first, is it ever possible to rule out direct influences in a psychology setting? This remains an open question, but we suspect the answer is negative. In physics one can always reproduce the results of quantum theory with a model which is non-contextual but non-local  \citep{Bohm1952}. In physics such accounts can be ruled out on the basis of a {\em physical} principle, locality, but  this is an additional assumption going beyond statements about the statistics of measurements. There is nothing equivalent in psychology that would supply such a clear cut limit on the set of allowable models. 

Second, if we cannot rule out models involving direct influence, does ruling out models involving the weaker notion of signalling tell us anything useful? In one sense the answer is clearly no - violations of contextual inequalities such as Eq.(\ref{eq1.5}) have been billed as tests of the necessity of a contextual (quantum) account of human decision making, and we have seen that such violations do not in fact rule out all possible non-contextual accounts, and therefore cannot definitively prove the necessity of a quantum model for such data. 

Does this mean contextual inequalities have nothing to teach us in psychology? Not necessarily. It has previously been argued \citep{Yearsley2014} that data satisfying Dzhafarov and Kujala's (2015) inequalities presents us with a choice - either we can construct a model which is non-contextual but which involves unobservable direct influences, or we can construct a model which only involves observed quantities, but which combines them in a contextual way. The correct way to proceed is not fixed by any mathematical law, but depends on the goals of the researcher.

Added note: after completion of this work we were made aware that a number of other authors have made closely related observations, including \cite{AF2019}, \cite{Cal2018} and \cite{Jones2019}. These connections will be addressed in future publications.

\section{Acknowledgments}
We are grateful to Samson Abramsky, Jerome Busemeyer, Ehtibar Dzhafarov,
Andrew Simmons and Rob Spekkens for useful communications on this topic.

\bibliography{Commentbibupdated}

\begin{thebibliography}{}

\bibitem [\protect \citeauthoryear {%
Abramsky%
\ \BBA {} Brandenburger%
}{%
Abramsky%
\ \BBA {} Brandenburger%
}{%
{\protect \APACyear {2011}}%
}]{%
Abramsky2011}
\APACinsertmetastar {%
Abramsky2011}%
\begin{APACrefauthors}%
Abramsky, S.%
\BCBT {}\ \BBA {} Brandenburger, A.%
\end{APACrefauthors}%
\unskip\
\newblock
\APACrefYearMonthDay{2011}{}{}.
\newblock
{\BBOQ}\APACrefatitle {The sheaf-theoretic structure of non-locality and
  contextuality.} {The sheaf-theoretic structure of non-locality and
  contextuality.}{\BBCQ}
\newblock
\APACjournalVolNumPages{New Journal of Physics}{}{13}{113036}.
\PrintBackRefs{\CurrentBib}

\bibitem [\protect \citeauthoryear {%
Aerts%
, Gabora%
\BCBL {}\ \BBA {} Sozzo%
}{%
Aerts%
\ \protect \BOthers {.}}{%
{\protect \APACyear {2013}}%
}]{%
Aerts2013}
\APACinsertmetastar {%
Aerts2013}%
\begin{APACrefauthors}%
Aerts, D.%
, Gabora, L.%
\BCBL {}\ \BBA {} Sozzo, S.%
\end{APACrefauthors}%
\unskip\
\newblock
\APACrefYearMonthDay{2013}{}{}.
\newblock
{\BBOQ}\APACrefatitle {Concepts and their dynamics: A quantum-theoretic
  modeling of human thought} {Concepts and their dynamics: A quantum-theoretic
  modeling of human thought}.{\BBCQ}
\newblock
\APACjournalVolNumPages{Topics in Cognitive Science}{5}{4}{737-772}.
\PrintBackRefs{\CurrentBib}

\bibitem [\protect \citeauthoryear {%
Asano%
, Hashimoto%
, Khrennikov%
, Ohya%
\BCBL {}\ \BBA {} Tanaka%
}{%
Asano%
\ \protect \BOthers {.}}{%
{\protect \APACyear {2014}}%
}]{%
Asano2014}
\APACinsertmetastar {%
Asano2014}%
\begin{APACrefauthors}%
Asano, M.%
, Hashimoto, T.%
, Khrennikov, A\BPBI Y.%
, Ohya, M.%
\BCBL {}\ \BBA {} Tanaka, Y.%
\end{APACrefauthors}%
\unskip\
\newblock
\APACrefYearMonthDay{2014}{}{}.
\newblock
{\BBOQ}\APACrefatitle {Violation of contextual generalization of the
  {L}eggett-{G}arg inequality for recognition of ambiguous figures} {Violation
  of contextual generalization of the {L}eggett-{G}arg inequality for
  recognition of ambiguous figures}.{\BBCQ}
\newblock
\APACjournalVolNumPages{Physica Scripta}{T163}{}{14006}.
\PrintBackRefs{\CurrentBib}

\bibitem [\protect \citeauthoryear {%
Atmanspacher%
\ \BBA {} Filk%
}{%
Atmanspacher%
\ \BBA {} Filk%
}{%
{\protect \APACyear {2019}}%
}]{%
AF2019}
\APACinsertmetastar {%
AF2019}%
\begin{APACrefauthors}%
Atmanspacher, H.%
\BCBT {}\ \BBA {} Filk, T.%
\end{APACrefauthors}%
\unskip\
\newblock
\APACrefYearMonthDay{2019}{}{}.
\newblock
{\BBOQ}\APACrefatitle {Contextuality revisited - Signaling may differ from
  communicating. In {A}. de {B}arros and {C}. {M}ontemayor (eds.), {Q}uanta and
  mind: Essays on the connection between quantum mechanics and the
  consciousness.} {Contextuality revisited - signaling may differ from
  communicating. in {A}. de {B}arros and {C}. {M}ontemayor (eds.), {Q}uanta and
  mind: Essays on the connection between quantum mechanics and the
  consciousness.}{\BBCQ}
\newblock
\APACjournalVolNumPages{Synthese Library.}{}{}{}.
\PrintBackRefs{\CurrentBib}

\bibitem [\protect \citeauthoryear {%
Basieva%
, Cervantes%
, Dzhafarov%
\BCBL {}\ \BBA {} Khrennikov%
}{%
Basieva%
\ \protect \BOthers {.}}{%
{\protect \APACyear {2019}}%
}]{%
Basieva2019}
\APACinsertmetastar {%
Basieva2019}%
\begin{APACrefauthors}%
Basieva, I.%
, Cervantes, V\BPBI H.%
, Dzhafarov, E\BPBI N.%
\BCBL {}\ \BBA {} Khrennikov, K.%
\end{APACrefauthors}%
\unskip\
\newblock
\APACrefYearMonthDay{2019}{}{}.
\newblock
{\BBOQ}\APACrefatitle {True Contextuality Beats Direct Influences in Human
  Decision Making} {True contextuality beats direct influences in human
  decision making}.{\BBCQ}
\newblock
\APACjournalVolNumPages{Accepted for publication in Journal of Experimental
  Psychology: General. Preprint available from
  https://arxiv.org/abs/1807.05684}{}{}{}.
\PrintBackRefs{\CurrentBib}

\bibitem [\protect \citeauthoryear {%
Bohm%
}{%
Bohm%
}{%
{\protect \APACyear {1952}}%
}]{%
Bohm1952}
\APACinsertmetastar {%
Bohm1952}%
\begin{APACrefauthors}%
Bohm, D.%
\end{APACrefauthors}%
\unskip\
\newblock
\APACrefYearMonthDay{1952}{}{}.
\newblock
{\BBOQ}\APACrefatitle {A Suggested Interpretation of the Quantum Theory in
  Terms of 'Hidden Variables' {I}} {A suggested interpretation of the quantum
  theory in terms of 'hidden variables' {I}}.{\BBCQ}
\newblock
\APACjournalVolNumPages{Physical Review}{85}{2}{166-179}.
\PrintBackRefs{\CurrentBib}

\bibitem [\protect \citeauthoryear {%
Bruza%
, Kitto%
, Ramm%
\BCBL {}\ \BBA {} Sitbon%
}{%
Bruza%
, Kitto%
\BCBL {}\ \protect \BOthers {.}}{%
{\protect \APACyear {2015}}%
}]{%
Bruza2015}
\APACinsertmetastar {%
Bruza2015}%
\begin{APACrefauthors}%
Bruza, P\BPBI D.%
, Kitto, K.%
, Ramm, B\BPBI J.%
\BCBL {}\ \BBA {} Sitbon, L.%
\end{APACrefauthors}%
\unskip\
\newblock
\APACrefYearMonthDay{2015}{}{}.
\newblock
{\BBOQ}\APACrefatitle {A probabilistic framework for analysing the
  compositionality of conceptual combinations} {A probabilistic framework for
  analysing the compositionality of conceptual combinations}.{\BBCQ}
\newblock
\APACjournalVolNumPages{Journal of Mathematical Psychology}{67}{}{26-38}.
\PrintBackRefs{\CurrentBib}

\bibitem [\protect \citeauthoryear {%
Bruza%
, Wang%
\BCBL {}\ \BBA {} Busemeyer%
}{%
Bruza%
, Wang%
\BCBL {}\ \BBA {} Busemeyer%
}{%
{\protect \APACyear {2015}}%
}]{%
Bruza2015b}
\APACinsertmetastar {%
Bruza2015b}%
\begin{APACrefauthors}%
Bruza, P\BPBI D.%
, Wang, Z.%
\BCBL {}\ \BBA {} Busemeyer, J\BPBI R.%
\end{APACrefauthors}%
\unskip\
\newblock
\APACrefYearMonthDay{2015}{}{}.
\newblock
{\BBOQ}\APACrefatitle {Quantum cognition: a new theoretical approach to
  psychology} {Quantum cognition: a new theoretical approach to
  psychology}.{\BBCQ}
\newblock
\APACjournalVolNumPages{Trends in Cognitive Sciences}{17}{7}{383-393}.
\PrintBackRefs{\CurrentBib}

\bibitem [\protect \citeauthoryear {%
Cavalcanti%
}{%
Cavalcanti%
}{%
{\protect \APACyear {2018}}%
}]{%
Cal2018}
\APACinsertmetastar {%
Cal2018}%
\begin{APACrefauthors}%
Cavalcanti, E\BPBI G.%
\end{APACrefauthors}%
\unskip\
\newblock
\APACrefYearMonthDay{2018}{}{}.
\newblock
{\BBOQ}\APACrefatitle {Classical causal models for Bell and {K}ochen-{S}pecker
  inequality violations require fine-tuning.} {Classical causal models for bell
  and {K}ochen-{S}pecker inequality violations require fine-tuning.}{\BBCQ}
\newblock
\APACjournalVolNumPages{Phys. Rev. X}{8}{}{021018}.
\PrintBackRefs{\CurrentBib}

\bibitem [\protect \citeauthoryear {%
Cervantes%
\ \BBA {} Dzhafarov%
}{%
Cervantes%
\ \BBA {} Dzhafarov%
}{%
{\protect \APACyear {2018}}%
}]{%
Cervantes2018}
\APACinsertmetastar {%
Cervantes2018}%
\begin{APACrefauthors}%
Cervantes, V\BPBI H.%
\BCBT {}\ \BBA {} Dzhafarov, E\BPBI N.%
\end{APACrefauthors}%
\unskip\
\newblock
\APACrefYearMonthDay{2018}{}{}.
\newblock
{\BBOQ}\APACrefatitle {Snow {Q}ueen is evil and beautiful: {E}xperimental
  evidence for probabilistic contextuality in human choices} {Snow {Q}ueen is
  evil and beautiful: {E}xperimental evidence for probabilistic contextuality
  in human choices}.{\BBCQ}
\newblock
\APACjournalVolNumPages{Decision}{5}{}{193-204}.
\PrintBackRefs{\CurrentBib}

\bibitem [\protect \citeauthoryear {%
Clemente%
\ \BBA {} Kofler%
}{%
Clemente%
\ \BBA {} Kofler%
}{%
{\protect \APACyear {2015}}%
}]{%
CK2015}
\APACinsertmetastar {%
CK2015}%
\begin{APACrefauthors}%
Clemente, L.%
\BCBT {}\ \BBA {} Kofler, J.%
\end{APACrefauthors}%
\unskip\
\newblock
\APACrefYearMonthDay{2015}{}{}.
\newblock
{\BBOQ}\APACrefatitle {Necessary and sufficient conditions for macroscopic
  realism from quantum mechanics.} {Necessary and sufficient conditions for
  macroscopic realism from quantum mechanics.}{\BBCQ}
\newblock
\APACjournalVolNumPages{Phys. Rev. A}{91}{}{062103}.
\PrintBackRefs{\CurrentBib}

\bibitem [\protect \citeauthoryear {%
Dzhafarov%
\ \BBA {} Kujala%
}{%
Dzhafarov%
\ \BBA {} Kujala%
}{%
{\protect \APACyear {2015}}%
}]{%
DK}
\APACinsertmetastar {%
DK}%
\begin{APACrefauthors}%
Dzhafarov, E\BPBI N.%
\BCBT {}\ \BBA {} Kujala, J\BPBI V.%
\end{APACrefauthors}%
\unskip\
\newblock
\APACrefYearMonthDay{2015}{}{}.
\newblock
{\BBOQ}\APACrefatitle {Generalizing {B}ell-type and {L}eggett-{G}arg-type
  Inequalities to Systems with Signaling} {Generalizing {B}ell-type and
  {L}eggett-{G}arg-type inequalities to systems with signaling}.{\BBCQ}
\newblock
\APACjournalVolNumPages{arXiv}{1407.2886v7}{}{}.
\PrintBackRefs{\CurrentBib}

\bibitem [\protect \citeauthoryear {%
Dzhafarov%
, Kujala%
, Cervantes%
, Zhang%
\BCBL {}\ \BBA {} Jones%
}{%
Dzhafarov%
, Kujala%
\BCBL {}\ \protect \BOthers {.}}{%
{\protect \APACyear {2016}}%
}]{%
Dzhafarov2016}
\APACinsertmetastar {%
Dzhafarov2016}%
\begin{APACrefauthors}%
Dzhafarov, E\BPBI N.%
, Kujala, J\BPBI V.%
, Cervantes, V\BPBI H.%
, Zhang, R.%
\BCBL {}\ \BBA {} Jones, M.%
\end{APACrefauthors}%
\unskip\
\newblock
\APACrefYearMonthDay{2016}{}{}.
\newblock
{\BBOQ}\APACrefatitle {On contextuality in behavioural data} {On contextuality
  in behavioural data}.{\BBCQ}
\newblock
\APACjournalVolNumPages{Philosophical Transactions of the Royal Society
  A}{374}{}{20150234}.
\PrintBackRefs{\CurrentBib}

\bibitem [\protect \citeauthoryear {%
Dzhafarov%
, Zhang%
\BCBL {}\ \BBA {} Kujala%
}{%
Dzhafarov%
, Zhang%
\BCBL {}\ \BBA {} Kujala%
}{%
{\protect \APACyear {2016}}%
}]{%
Dzhafarov2015}
\APACinsertmetastar {%
Dzhafarov2015}%
\begin{APACrefauthors}%
Dzhafarov, E\BPBI N.%
, Zhang, R.%
\BCBL {}\ \BBA {} Kujala, J\BPBI V.%
\end{APACrefauthors}%
\unskip\
\newblock
\APACrefYearMonthDay{2016}{}{}.
\newblock
{\BBOQ}\APACrefatitle {Is there contextuality in behavioural and social
  systems?} {Is there contextuality in behavioural and social systems?}{\BBCQ}
\newblock
\APACjournalVolNumPages{Philosophical Transactions of the Royal Society
  A}{374}{}{20150099}.
\PrintBackRefs{\CurrentBib}

\bibitem [\protect \citeauthoryear {%
Fine%
}{%
Fine%
}{%
{\protect \APACyear {1982}}%
}]{%
Fine1982}
\APACinsertmetastar {%
Fine1982}%
\begin{APACrefauthors}%
Fine, A.%
\end{APACrefauthors}%
\unskip\
\newblock
\APACrefYearMonthDay{1982}{}{}.
\newblock
{\BBOQ}\APACrefatitle {Hidden Variables, Joint Probability, and the {B}ell
  Inequalities} {Hidden variables, joint probability, and the {B}ell
  inequalities}.{\BBCQ}
\newblock
\APACjournalVolNumPages{Phys. Rev. Lett.}{48}{}{291--295}.
\PrintBackRefs{\CurrentBib}

\bibitem [\protect \citeauthoryear {%
Jones%
}{%
Jones%
}{%
{\protect \APACyear {2019}}%
}]{%
Jones2019}
\APACinsertmetastar {%
Jones2019}%
\begin{APACrefauthors}%
Jones, M.%
\end{APACrefauthors}%
\unskip\
\newblock
\APACrefYearMonthDay{2019}{}{}.
\newblock
{\BBOQ}\APACrefatitle {Relating causal and probabilistic approaches to
  contextuality.} {Relating causal and probabilistic approaches to
  contextuality.}{\BBCQ}
\newblock
\APACjournalVolNumPages{arxiv 1906.0271}{}{}{}.
\PrintBackRefs{\CurrentBib}

\bibitem [\protect \citeauthoryear {%
Oaksford%
\ \BBA {} Chater%
}{%
Oaksford%
\ \BBA {} Chater%
}{%
{\protect \APACyear {2007}}%
}]{%
Oaksford2007}
\APACinsertmetastar {%
Oaksford2007}%
\begin{APACrefauthors}%
Oaksford, M.%
\BCBT {}\ \BBA {} Chater, N.%
\end{APACrefauthors}%
\unskip\
\newblock
\APACrefYear{2007}.
\newblock
\APACrefbtitle {Bayesian {R}ationality: {T}he Probabilistic Approach to Human
  Reasoning.} {Bayesian {R}ationality: {T}he probabilistic approach to human
  reasoning.}
\newblock
\APACaddressPublisher{}{OUP}.
\PrintBackRefs{\CurrentBib}

\bibitem [\protect \citeauthoryear {%
Yearsley%
\ \BBA {} Pothos%
}{%
Yearsley%
\ \BBA {} Pothos%
}{%
{\protect \APACyear {2014}}%
}]{%
Yearsley2014}
\APACinsertmetastar {%
Yearsley2014}%
\begin{APACrefauthors}%
Yearsley, J\BPBI M.%
\BCBT {}\ \BBA {} Pothos, E\BPBI M.%
\end{APACrefauthors}%
\unskip\
\newblock
\APACrefYearMonthDay{2014}{}{}.
\newblock
{\BBOQ}\APACrefatitle {Challenging the classical notion of time in cognition: a
  quantum perspective} {Challenging the classical notion of time in cognition:
  a quantum perspective}.{\BBCQ}
\newblock
\APACjournalVolNumPages{Proceedings of the Royal Society
  B}{281}{1781}{20133056}.
\PrintBackRefs{\CurrentBib}

\end{thebibliography}

\end{document}